\documentclass[floatfix, twocolumn, prb]{revtex4}

\usepackage{color}
\usepackage{multirow}
\usepackage{epsfig}
\usepackage{graphicx}
\usepackage{amsmath}
\usepackage{amssymb}
\usepackage{bm}
\usepackage{dcolumn}
\usepackage{braket}
\usepackage{bbold}
\usepackage{ulem}
\usepackage{soul}
\usepackage[dvipsnames]{xcolor}

\begin{document}

\title{Effective $J\!=\!1/2$ insulating state in Ruddlesden-Popper iridates:\\
An LDA+DMFT study}

\author{Hongbin~Zhang}
\email[corresp.\ author: ]{hzhang@physics.rutgers.edu}
\author{Kristjan Haule}
\author{David Vanderbilt}
\affiliation{Department of Physics and Astronomy, Rutgers University, Piscataway, USA}

\date{today}

\begin{abstract}
Using {\it ab-initio} methods for correlated electrons in solids, we investigate
the metal-insulator transition across the Ruddlesden-Popper (RP) series of
iridates and explore the robustness of the $J_\text{eff}\!=\!1/2$ state against 
band effects due to itineracy, tetragonal distortion, octahedral
rotation and Coulomb interaction. We predict the effects of epitaxial
strain on the optical conductivity, magnetic moments, and 
$J_\text{eff}\!=\!1/2$ ground-state wave functions in the RP series.  
To describe the solution of the
many-body problem in an intuitive picture, we introduce a concept of
energy-dependent atomic states, which strongly resemble the atomic $J_\text{eff}\!=\!1/2$
states but with coefficients that are energy/time-dependent.  We
demonstrate that the deviation from the ideal $J_\text{eff}\!=\!1/2$ state is
negligible at short time scales for both single- and double-layer
iridates, while it becomes quite significant for Sr$_3$Ir$_2$O$_7$ at long
times and low energy. Interestingly, Sr$_2$IrO$_4$ is positioned
very close to the $SU(2)$ limit, with only $\sim\!3\%$ deviation from the
ideal $J_\text{eff}\!=\!1/2$ situation.
\end{abstract}

\maketitle

Metal-insulator transitions are very common in $3d$ transition-metal
oxides due to the small bandwidth of the $3d$ orbitals and the poorly
screened electron-electron interaction on the $3d$
ions.~\cite{Imada:1998}  Due to the larger spatial extent of
the $5d$ orbitals, the $5d$ transition-metal oxides are expected
to be more itinerant.  However, because of strong spin-orbit coupling (SOC),
many $5d$ transition-metal oxides show significant
electron-electron correlations and even metal-insulator transitions of
possible Mott type.
One of the most well-studied $5d$ systems at the
localization-delocalization boundary is the Ruddlesden-Popper (RP)
series of iridates with chemical formula
Sr$_{n+1}$Ir$_n$O$_{3n+1}$, where $n$ is the number of
SrIrO$_3$ perovskite layers sandwiched between extra SrO layers. 
Experimentally, Sr$_2$IrO$_4$ (214) is a Mott-like magnetic
insulator with canted in-plane anti-ferromagnetic (AFM)
ordering,~\cite{Cao:1998, Fujiyama:2012} Sr$_3$Ir$_2$O$_7$ (327) is
a narrow-gap Slater-like
AFM insulator with moments aligned along the
$c$-axis,~\cite{Cao:2002, Fujiyama:2012a} while SrIrO$_3$ (113) is a
correlated metal.~\cite{Cao:2007}
These compounds have attracted tremendous attention
recently~\cite{Kim:2008, Jackeli:2009, Wang:2011} because of
the similarity between 214 and the parent compound La$_2$CuO$_4$ of the cuprate
superconductors: the structures are the same,
the low-energy properties can be modeled by a single band Hubbard-type
model,~\cite{Kim:2008} and the magnetic spin wave spectrum in Sr$_2$IrO$_4$ 
shows no observable spin-wave gap.~\cite{Kim:2012a}
This is quite unexpected because the large SOC and
IrO$_6$ octahedral rotations lead to significant Dzyaloshinsky-Moria
interaction.  On the other hand, a very different spin-wave spectrum with a large magnon
gap was found in the double-layer 327 compound.~\cite{Kim:2012b}

Kim {\it et al.}~\cite{Kim:2008} proposed that the strong SOC
and the large octahedral crystal-field splitting between the
$t_{2g}$ and $e_{g}$ states produce an effective $J_\text{eff}\!=\!1/2$ state
on the Ir$^{4+}$ ion, where the magnetic moment is isotropic and $SU(2)$-invariant.
The $J_\text{eff}\!=\!1/2$ states form a Kramers-doublet and
contain an equal mixture of $d_{xz}$, $d_{yz}$ and $d_{xy}$ orbitals in the
form
$\ket{\psi_{-1/2}}\!=\!\left( \ket{d_{xy}\!\uparrow} + \ket{d_{yz}\!\downarrow}
   +i\ket{d_{xz}\!\downarrow}\right)/\sqrt{3}$
and
$\ket{\psi_{+1/2}}\!=\!\left( -\ket{d_{xy}\!\downarrow} +\ket{d_{yz}\!\uparrow}
-i\ket{d_{xz}\uparrow}\right)/\sqrt{3}$.
This is the ground state of a single ion
in a cubic crystal environment, which carries a magnetic moment of
1$\,\mu_B$. Experimentally, the Ir ions in 214 have
significantly smaller sublattice magnetic moments of the order of
$0.5\,\mu_B$,~\cite{Cao:1998} which demonstrates the importance of
itineracy in this system. Moreover, the tetragonal distortions in all
members of the RP series of iridates are large, leading to a crystal-field splitting
$\Lambda\!\equiv\!\varepsilon_{xz}-\varepsilon_{xy}$
between $d_{xy}$ and $d_{xz/yz}$ orbitals that breaks the $SU(2)$
invariance of the magnetic moments. Nevertheless, the absence of
resonant X-ray magnetic scattering intensity at the L$_2$ edge suggests
that the electronic state in both the single-layer~\cite{Kim:2009}
and double-layer~\cite{Boseggia:2012}
members of the RP series are quite close to the $J_\text{eff}\!=\!1/2$ limit.

In this work we have performed dynamical mean field theory (DMFT)
calculations~\cite{Kotliar:review} in a charge self-consistent implementation on top of
density functional theory (DFT), allowing for a realistic treatment
of competing crystal-field, spin-orbit, and Coulomb interactions.
Our work provides quantitative answers to the
questions of how good is the $J_\text{eff}\!=\!1/2$ description in
the RP series of iridates, and how large is the deviation from the
isotropic $SU(2)$ magnetic response.  
We show that the answers to these questions have a time-scale or energy-scale dependence, thus resolving some points of controversy in the literature.
%

Our {\it ab-initio} DFT+DMFT implementation\cite{Haule:2010}
is based on the WIEN2k package.~\cite{wien2k}
The impurity model was solved using the continuous-time quantum Monte
Carlo method.\cite{Haule:CTQMC}  To minimize the sign
problem, a proper choice of the local basis is required. We first
chose pseudo-cubic local axes on each Ir atom adapted to the IrO$_6$
octahedra, and then performed an exact diagonalization of the
impurity levels.
The calculations were fully self-consistent in the charge density, chemical
potential, impurity levels, lattice and impurity Green's
functions, hybridizations, and self-energies.
The energy range in computing the hybridizations and self-energies spanned
a $20\,$eV window around the Fermi energy (E$_\text{F}$), allowing us to use
a set of system-independent local Coulomb interaction parameters
$U$ and $J$.  This is in contrast to other DMFT calculations on
iridates~\cite{Martins:2011,Arita:2012} where downfolding to Ir
$t_{2g}-$orbitals was performed, so that the proper values of $U$ and $J$ depend
sensitively on the screening by the bands eliminated from the model.
The value of $U$ for our large energy window was estimated for
the undistorted 214 with tetragonal structure (P4/mmm) using the method of
Ref.~\onlinecite{Kutepov:U}, which leads to $U\!\approx\!4.5\,$eV and
$J\!\approx\!0.8\,$eV.
To properly simulate the non-collinear magnetic state in the 214 compound,
we chose different local coordinates on each Ir atom, with local
quantization axis of spin in the direction of the ordered magnetic moment, 
and used proper Wigner rotations of spins and orbitals to transform the
local self-energy to a common global axis.
The DOS and optical conductivities were computed from analytic
continuation of the self-energy from the imaginary frequency axis to
real frequencies using an auxiliary Green's function and the maximum-entropy
method.
The DMFT calculations were performed at 50$\,$K, below the AFM
transition temperature of the 214 ($240\,$K)~\cite{Cao:1998} and
327 ($280\,$K)~\cite{Cao:2002}.
For comparison, we also carried out DFT+U calculations using the
full-potential linearized augmented plane wave (FLAPW) method as
implemented in the Elk code.~\cite{elk} Since DFT+U does not include
screening effects, the fully screened interaction on Ir is needed,
which is here determined to be $U\!\approx\!2.5\,$eV by
fitting to the DMFT results.
VASP was used to relax the structures when epitaxial strain is
considered, with atomic forces converged to $10^{-3}\,\textrm{eV}/\textrm{\AA}$ in the
GGA approximation.~\cite{vasp}

\begin{figure}[ht]
\includegraphics[width=8.0cm]{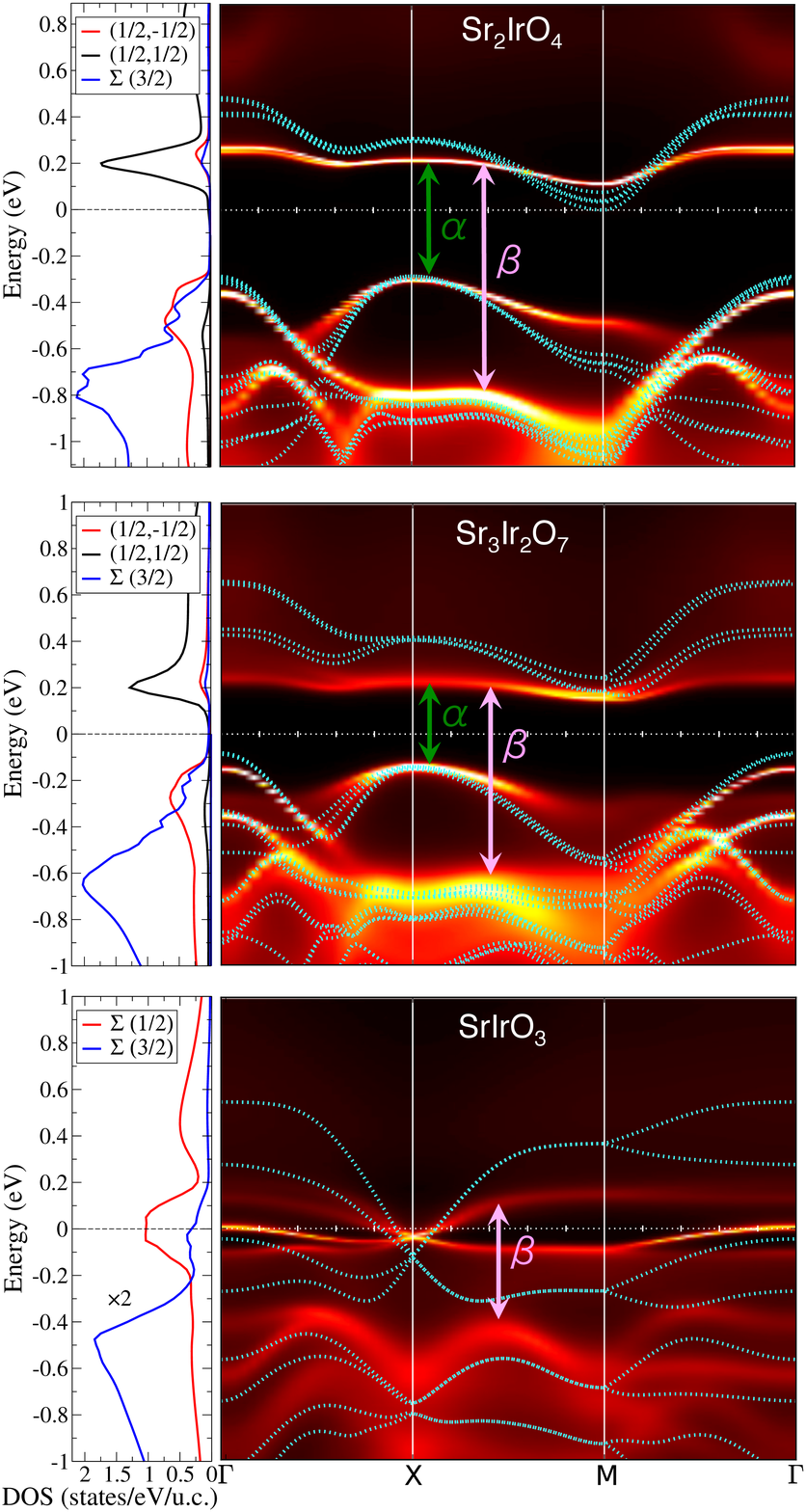}
\caption{(Color online)
Spectral functions and orbital-resolved densities of states (DOS)
  obtained by the DMFT method for the RP series of iridates.
  Arrows indicate optical transitions corresponding to the peaks
  in the optical conductivities shown in Fig.~\ref{fig:optics}. Dotted
  lines denote the band structures obtained by GGA+U calculations.
  For 214 and 327, the GGA+U band structures and the DMFT
  spectral functions are aligned by fixing the position of the topmost
  valence state at $X$. Orbitals $(1/2,\pm 1/2)$ correspond to
  $\psi_{\pm 1/2}$, and $\sum(3/2)$ stands for the sum over the
  remaining $t_{2g}$ states, {\it i.e.}, $J_\text{eff}\!=\!3/2$ states.
}
\label{fig:akw}
\end{figure}
The resulted DMFT spectral functions are presented in Fig.~\ref{fig:akw}
for the 214, 327, and 113 compounds, with color coding showing the
spectral intensity along $\Gamma\!-\!X\!-\!M\!-\!\Gamma$,
and the side panels displaying the corresponding orbital-resolved DOS. 
The dotted lines show an overlay of the band structures from the GGA+U calculations.
Within DMFT, the insulating gap in 214 and 327 are approximately $400$ and
$300$\,meV, respectively. There is a significant amount of incoherent
spectral weight in the gap, which shows up in the DOS, 
but is hardly noticeable in the spectral-function plot.  
The unoccupied electronic states are of mainly $J_\text{eff}\!=\!1/2$ character. 
The first valence state at $X$ is also of $J_\text{eff}\!=\!1/2$ character, while the
first valence state at $\Gamma$ is of $J_\text{eff}\!=\!3/2$ character, hence the
occupied states are an equal mixture of $J_\text{eff}\!=\!1/2$ and $J_\text{eff}\!=\!3/2$ states. 
In the DMFT calculation, the topmost valence state at $X$ is
about $40\,$meV closer to E$_\text{F}$ than the first valence state at the $\Gamma$ point, in
agreement with recent ARPES measurements,~\cite{Wojek:2012,Wang:2013}
but in contrast to GGA+U results, where the first
valence state is at the $\Gamma$ point. This is because the filled
orbitals (here $J_\text{eff}\!=\!3/2$) tend to be repelled from the Fermi level in DMFT. 
In 327, the $J_\text{eff}\!=\!3/2$ tail at
$\Gamma$ is split into two peaks due to its double-layer structure,
consistent with experimental results.~\cite{King:2013}
Finally, the 113 compound is a strongly correlated
metal,~\cite{Cao:2007} with very flat bands around E$_\text{F}$. 
The effective mass of the $t_{2g}$ states is quite large, 
{\it e.g.,} the effective mass of the hole pocket around $\Gamma$
is about nine times that of bare electron.

A closer look at the orbital-resolved DOS in Fig.~\ref{fig:akw} reveals
that the $J_\text{eff}\!=\!1/2$ states are not fully polarized.  This
is due to significant itineracy effects and hybridization between
Ir~$5d$ and O~$2p$ states.  The resulting occupation number
for the occupied $J_\text{eff}\!=\!1/2$ orbitals is about 0.65 (0.45)
in the 214 (327) compound.  Consequently, this leads to substantially
reduced magnetic moments compared to the ideal $J_\text{eff}\!=\!1/2$ value of
$1\,\mu_B$, down to about $0.55\,\mu_B$ (214) and $0.58\,\mu_B$ (327).

While the most part the GGA+U band structures are in fairly
good agreement with the DMFT spectral functions (Fig.~\ref{fig:akw}),
there are also some differences.
The band gaps in the 214 and 327 compounds within GGA+U
are of almost equal size, about $270\,$meV ($U_\text{GGA+U}\!=\!2.5\,$eV),
while there is a clear gap reduction of about $100\,$ meV in the DMFT
calculations.  This is not surprising, given that the fully screened $U$,
required by GGA+U, should be reduced in the more itinerant 327
compounds. For 113, the Fermi surface in DMFT is quite similar to the
GGA Fermi surface (not shown), but the bandwidth is strongly renormalized.
This is not the case in GGA+U, where the
hole pocket at $\Gamma$ is missing.  Recent ARPES
measurements~\cite{Nie} confirmed the existence of the hole pocket at
$\Gamma$ with a strongly enhanced effective mass, in agreement with the
DMFT results.

We mention in passing that the paramagnetic calculation for 214 
is not insulating in DFT+DMFT, but a very bad metal, in
agreement with the DMFT calculations of Ref.~\onlinecite{Arita:2012}, but in
disagreement with Ref.~\onlinecite{Martins:2011}. 
The single-site DMFT calculations describe exactly the correlations local to the Ir
sites, and also the correlations of infinite range with AFM ordering.
Based on magnetic x-ray scattering, Fujiyama~{\it et al.}~\cite{Fujiyama:2012}
reported a large but finite correlation length exceeding 100
lattice spacings even 20$\,$K above the N\'{e}el temperature
in the 214 compound.
The ``marginal Mott insulator,''~\cite{Fujiyama:2012} a term coined
to describe such a short-range ordered state, is not captured by
the single site DMFT method, but requires cluster extensions, 
so unfortunately our modeling cannot account for
the properties of this compound near its AFM phase transition.

\begin{figure}
\includegraphics[width=8.5cm]{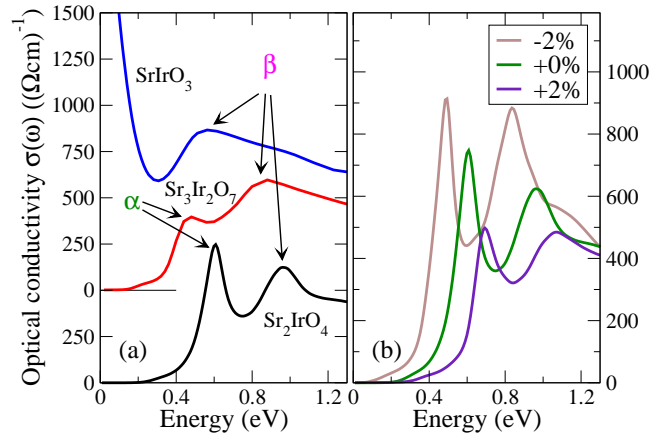}
\caption{(Color online)
a) Computed optical conductivities $\sigma_1$ for three iridates
at their experimental lattice constants. For clarity, $\sigma_1$
is shifted up by 500$\,(\Omega\textrm{cm})^{-1}$ for Sr$_3$Ir$_2$O$_7$
and SrIrO$_3$;
$\alpha$ and $\beta$ indicate peak positions corresponding to the
vertical transitions marked in Fig.~\ref{fig:akw}.
b) Dependence of $\sigma_1$ on epitaxial strain ($0\%$ and $\pm2\%$) in
Sr$_2$IrO$_4$.
}
\label{fig:optics}
\end{figure}

The optical conductivities for the three compounds obtained by the DMFT
calculations are shown in Fig.~\ref{fig:optics}. In agreement with
experiments,~\cite{Moon:2008} the optical conductivities of the 214 and
327 insulating compounds have two peaks denoted by $\alpha$ and
$\beta$ in the low-energy range ($0\!\sim\!1.2\,$eV).
As proposed in Ref.~\onlinecite{Moon:2009}, the $\alpha$ peak is
mainly due to the excitations from the highest valence band to the first
conduction band, both being primarily of $J_\text{eff}\!=\!1/2$ character.
The second peak is mainly due to the excitations from the lower valence
bands, of primarily $J_\text{eff}\!=\!3/2$ character, to the
conduction band of $J_\text{eff}\!=\!1/2$ character.  It is
interesting to note that the $J_\text{eff}\!=\!1/2$ and
$J_\text{eff}\!=\!3/2$ valence states strongly overlap in energy, as
is clear from the DOS (Fig.~\ref{fig:akw}).  Thus, there is no clear
separation between the two types of excitations in local
quantities, in disagreement with the simple cartoon of
Ref.~\onlinecite{Moon:2009}.  Nevertheless, the vertical excitations
probed by optics do give rise to two well separated peaks.  Going from
214 to 327, the $\alpha$ peak shifts by about $100\,$meV to 
lower energy, while it is replaced by a narrow Drude peak in the 113 compound.  The
$\beta$ peak broadens and shifts from $1.0\,$eV in 214 to $0.5\,$eV in
113, in good agreement with experiments.~\cite{Moon:2008} We notice
that there is an observable tail of optical conductivity in both
insulating compounds at low energy (spanning the region from
0.25$\,$eV to about 0.4$\,$eV in 214), which can be attributed to the
incoherent spectral weight in the gap.  A similar tail was also
found in optical experiments.\cite{Moon:2008, Moon:2009}

To shed more light on the nature of the $J_\text{eff}\!=\!1/2$
insulating state, we performed calculations for the RP series of iridates
under epitaxial strain, where the out-of-plane lattice parameter and
internal lattice coordinates were relaxed in the GGA approximation.
Fig.~\ref{fig:optics}(b) shows the optical conductivity for the 214 compound with 
$-2$\% (compressive), $0$\%, and $+2$\% (tensile) strain. Compressive (tensile)
strain substantially reduces (increases) the gap size due to an increase
(decrease) of the dominant in-plane hoppings. The two peaks are shifted to 
lower (higher) energy, and become sharper (broadened), while the overall
conductivity increases (decreases) under compressive (tensile) strain.
This is in good agreement with recent optical measurements on strained
214 thin films.~\cite{Nichols:2013}

\begin{figure}
\includegraphics[width=8.0cm]{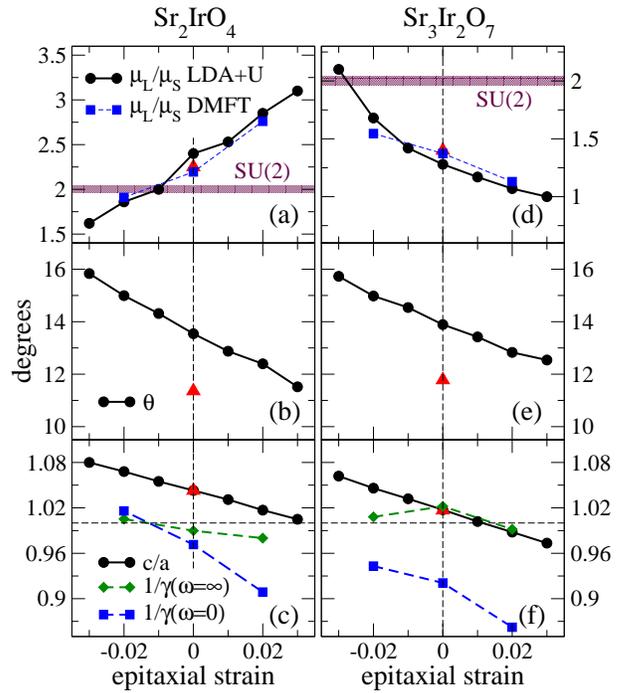}
\caption{ (colored online) 
Dependence of computed properties of 214 (a-c) and 327 (d-f) 
compounds on epitaxial strain in the range of $\pm3\%$.  
(a) and (d): orbital-to-spin moment ratio $\mu_\text{L}/\mu_\text{S}$.
(b) and (e): staggered rotation angle $\theta$ of IrO$_6$ octahedra.
(c) and (f): ratio of Ir-O bond lengths along $z$ and in-plane $c/a$,
and inverse of parameter $\gamma$ describing
the generalized $J_\text{eff}\!=\!1/2$ wave function (see main text).
Filled symbols with dashed (solid black) lines denote the results obtained 
by DMFT (GGA+U) calculations. Red triangles indicate GGA+U
results computed with unrelaxed experimental lattice parameters.}
\label{fig:214-strain}
\end{figure}

The evolution of the structural parameters under epitaxial strain is shown
in Fig.~\ref{fig:214-strain}.
Both the $c/a$ ratio, panels (c) and (f),  and 
the rotation angle $\theta$ of the IrO$_6$ octahedra, panels (b) and (e),
decrease nearly linearly with increasing in-plane lattice constants,
making the crystal structure less distorted.  These structural parameters are
in good agreement with recent experiments on thin films.~\cite{Nichols:2013}

For the ideal $J_\text{eff}\!=\!1/2$ state, which is $SU(2)$ invariant, the
orbital magnetic moment on the Ir is twice as large as the spin moment,
with values of $2/3$ and $1/3$ $\mu_B$, respectively.
Fig.~\ref{fig:214-strain}(a) and (d) show the $\mu_\text{L}/\mu_\text{S}$ ratio
as obtained by the DMFT and GGA+U
methods for the 214 and 327 compounds.
In 214, this ratio increases with tensile
strain, while it decreases in 327.
In the absence of strain, $\mu_\text{L}/\mu_\text{S}\!\approx\! 2.2$
for 214 and $\mu_\text{L}/\mu_\text{S}\!\approx\! 1.3$, which
demonstrates that the 214 compound has only a slightly larger orbital moment
than expected for the ideal $SU(2)$ situation, while 327 has a substantially
smaller orbital moment. For 214 (327) compounds, the $SU(2)$ point can
be reached by 1\% (3\%) compressive strain.
It is interesting to note that at this $SU(2)$ point, the IrO$_6$
octahedra are significantly elongated in the $z$-direction
($c/a\!\approx\!1.05$). 
Thus, the deviation from $SU(2)$ behavior is not
simply associated with tetragonality, as might have been expected.

To gain some understanding into this puzzling behavior of the magnetic
moments, we analyzed the DMFT hybridization function $\Delta(\omega)$,
which carries all the information about the crystal environment for an
electron on a given iridium site. It is defined by
  $1/(\omega-\Delta-\Sigma)\!=\!\sum_k \hat{P}\,  /(\omega+\mu-\varepsilon_k-\hat{P}^{-1}\Sigma)$.
Here $\Sigma$ is the
DMFT self-energy, $\varepsilon_k$ are the Kohn-Sham-like eigenvalues,
and $\hat{P}$ ($\hat{P}^{-1}$) is the projector (embedder) on the
Ir site.  In the high-frequency limit,
$\Delta(\omega\!\rightarrow\!\infty)\!=\!\sum_k\hat{P}\;(\varepsilon_k-\mu)$
is a matrix whose elements denote the atomic on-site energy levels.  It
includes both the crystal-field and SOC terms, and is directly
related to the so called single-ion anisotropy.
The low-energy counterpart $\Delta(\omega\!=\!0)$ is related to the
low-energy excitations such as spin waves.

For both 214 and
327 compounds, the hybridization function can be well represented by
the matrix
\begin{equation}
\Delta(\omega)\!=\!\left(\begin{array}{c|ccc}
 & xz\uparrow & yz\uparrow & xy\downarrow\\
\hline
xz\uparrow & \epsilon                    &     -i\frac{\lambda}{2}         &    i(\frac{\lambda}{2}+\delta) \\
yz\uparrow & i\frac{\lambda}{2}   & \epsilon                               &    -(\frac{\lambda}{2}+\delta) \\
xy\downarrow & -i(\frac{\lambda}{2}+\delta) & -(\frac{\lambda}{2}+\delta)  & \epsilon-\Lambda
\end{array}\right),
\label{Hybridization}
\end{equation}
where $\epsilon$ indicates the on-site energy, $\lambda$ the
SOC strength, $\Lambda$ the $d_{xz/yz}/d_{xy}$ crystal-field
splitting, and $\delta$ the renormalization of the SOC between
$d_{xy}$ and $d_{xz/yz}$ orbitals. 
In general, all of these parameters are frequency-dependent. 
For the above matrix, the largest eigenvalue corresponds to eigenvectors
\begin{equation}
\begin{split}
\ket{\psi_{+\frac{1}{2}}} &\!=\!-\sqrt{\frac{3-2\gamma(\omega)^2}{3}} \ket{d_{xy}\!\downarrow}
   + \frac{\gamma(\omega)}{\sqrt{3}}(\ket{d_{yz}\!\uparrow}-i\ket{d_{xz}\!\uparrow}), \\
\ket{\psi_{-\frac{1}{2}}} &\!=\! \sqrt{\frac{3-2\gamma(\omega)^2}{3}} \ket{d_{xy}\!\uparrow}
   + \frac{\gamma(\omega)}{\sqrt{3}}(\ket{d_{yz}\!\downarrow}+i\ket{d_{xz}\!\downarrow}).\\
 \end{split}
\end{equation}
This is a generalization of the ideal $J_\text{eff}\!=\!1/2$ wave function,
which is recovered when $\gamma\!=\!1$.
For small deviation from the ideal $SU(2)$ case ($\Lambda/\lambda\ll 1$
and $\delta/\lambda\ll 1$) one finds
$\gamma\!=\!1+\frac{2}{9}\frac{\Lambda-\delta}{\lambda}+\cdots\!=\!
1+\tilde{\gamma}$, where $\tilde{\gamma}\!\equiv\!\gamma-1$ is a small deviation of positive
or negative sign for $J_\text{eff}\!=\!1/2$ states that are
respectively expanded or contracted in the $z$-direction.
Within the low-energy subspace of $\psi_{\pm 1/2}$, the orbital and
spin moments along $z$ become
$\braket{\mu_\text{L}^z}\! =\!\frac{2}{3}\gamma^2\Delta n\!\approx\!(\frac{2}{3}+\frac{4}{3}\,\tilde{\gamma})\Delta n$ and
$\braket{\mu_\text{S}^z}\! =\!(\frac{4}{3}\gamma^2-1) \Delta n\!\approx\!(\frac{1}{3}+\frac{8}{3}\tilde{\gamma})\Delta n$,
while in-plane they are
$\braket{\mu_\text{L}^{xy}}\!=\!\frac{2}{3}\gamma\sqrt{3-2\gamma^2}\Delta n\!\approx\! (\frac{2}{3}-\frac{2}{3}\tilde{\gamma})\Delta n$ and
$\braket{\mu_\text{S}^{xy}}\!=\frac{3-2\gamma^2}{3} \Delta n\!\approx\!(\frac{1}{3}-\frac{4}{3}\tilde{\gamma})\Delta n$,
where $\Delta n\!\equiv\! n_{1/2}-n_{-1/2}$ is the difference of the occupation numbers of  the $\psi_{\pm1/2}$ states.
We checked that the {\it ab-initio}-obtained values for both orbital and
spin moments are well accounted for by these simplified expressions
in the low-energy limit, i.e., when $\gamma(\omega\!=\!0)$ is used.

The values of $1/\gamma(\omega\!=\!\infty)$ and
$1/\gamma(\omega\!=\!0)$, as obtained by exact diagonalization of the
hybridization matrix, are plotted in
Figs.~\ref{fig:214-strain}(c) and (f).
In a simplified model, where only splitting between $d_{xy}$ and $d_{yz/xz}$ due to
the tetragonal distortion of IrO$_6$ octahedra is considered, $1/\gamma$ is
expected to be directly proportional to the $c/a$ ratio and $1/\gamma\!=\!1.0$ 
when $c/a\!=\! 1$. Clearly this is not the case as shown in Figs.~\ref{fig:214-strain}(c) and (f).
In the high-frequency limit, $\gamma(\omega\!=\!\infty)$ deviates from the
ideal value of unity by less than 2\%, which shows that the single-ion
anisotropy is small in both compounds.  In the low-frequency limit,
which is more relevant for spin dynamics, we can approximate
$\gamma(\omega\!=\!0)\!\approx\!1.03 + 3.5 r$ in the 214 compound, where
$r$ denotes the amount of epitaxial strain.  In the absence of strain,
this is quite close to unity, and reaches the ideal value upon 0.9\%
compressive strain ($r\!=\!-0.009$).  The behavior in the 327 compound
is quite different,
where we might approximate $\gamma(\omega\!=\!0)\!\approx\!
1.08+2.5 r $.  A substantial deviation of 8\% from the $SU(2)$ value is
obtained in the absence of strain, and more than $3\%,$ compressive
strain is necessary to recover an isotropic magnetic moment.

Our {\it ab-initio} values of $\mu_\text{L}/\mu_\text{S}$ are
well accounted for by this simplified model when $\gamma(\omega\!=\!0)$ is
used to evaluate the magnetic moments.
In the 214 compound the
magnetic moments are ordered in-plane, so that
$\mu_\text{L}^{xy}/\mu_\text{S}^{xy}\!\approx\!
(\frac{2}{3}-\frac{2}{3}\tilde{\gamma})/(\frac{1}{3}-\frac{4}{3}\tilde{\gamma})\!\approx\!
2+6\tilde{\gamma}\!\approx\!2.2 + 21 r$, which is quite close to the
{\it ab-initio}-calculated line
in Fig.~\ref{fig:214-strain}(a). 
In the 327 compound, the AFM moments are
ordered in the $z$-direction, hence
$\mu_\text{L}^z/\mu_\text{S}^z\!\approx\!
(\frac{2}{3}+\frac{4}{3}\tilde{\gamma})/(\frac{1}{3}+\frac{8}{3}\tilde{\gamma})\!\approx\!
2-12 \tilde{\gamma}\!\approx\! 1.04 - 30 r$.
Due to the much larger deviation
$\tilde{\gamma}$, this approximation is not very accurate, but
it nevertheless gives a decreasing moment ratio
$\mu_\text{L}/\mu_\text{S}$ with tensile strain and a large
deviation of this ratio from the ideal $SU(2)$ value of 2.
Therefore, we conclude that the low-energy magnetic excitations in 214 are 
quite close to the ideal $SU(2)$ isotropic case, with only $3\%$ larger 
$d_{xz/yz}$ contributions to the ground-state wave function than
that of its $d_{xy}$ component. 
On the other hand, in 327 there is substantially
more $d_{xz/yz}$ character ($\approx\! 8\%$) in the low-energy wave
function, which leads to larger moments in the $z$-direction and a smaller
$\mu_\text{L}/\mu_\text{S}$ ratio.

Finally, let us comment on the {\it ab-initio} values of the entries
in the hybridization matrix, Eq.~\ref{Hybridization}. At high
frequency $\Delta(\omega\!\rightarrow\!\infty)$, the crystal-field
splitting is typically only $\Lambda\!\approx\!25\,$meV, while the SOC
strength $\lambda/2\!\approx\!250\,$meV, and the SOC enhancement
$\delta$ vanishes. This gives
$\gamma(\omega\!=\!\infty)\!\approx\!1.01$, as shown in
Figs.~\ref{fig:214-strain}(d) and (h). At low energy, the crystal-field
splitting is strongly enhanced by the hybridization effects of the
tetragonal crystal structure.  In 214, it typically takes a value of
$\Lambda\!\approx\!  140\,$meV. Unexpectedly, $\delta$ 
is of similar magnitude in the 214 compounds, hence
$\gamma(\omega\!=\!0)\!\approx\!1+ \frac{2}{9}(\Lambda-\delta)/\lambda$
is again close to unity.  In 327 the enhancement of crystal fields
is even larger ($\Lambda\!\approx\!300\,$meV), but $\delta$ is
somewhat smaller ($\delta\!\approx\!120\,$meV) compared to that in
214, so that a significant deviation from the ideal value is observed
($\gamma\!\approx\! 1.08$).  We note that the proximity to the
$SU(2)$-invariant point leads to 
isotropic moments and almost gapless spin wave excitations,
which is in
qualitative agreement with recent measurements where a small spin gap
was found in the 214 compound~\cite{Kim:2012a} and a large spin gap in
the 327 compound.~\cite{Kim:2012b} We also note that fast experimental
probes should be more consistent with our
$\Delta(\omega\!\rightarrow\!\infty)$ limit, with a more $SU(2)-$like
symmetric response ($\gamma(\omega\!=\!\infty)\!\approx\! 1$), while
low-energy probes should see larger deviations from this limit
($\gamma(\omega\!=\!0)$ as in Fig.~\ref{fig:214-strain}).

{\bf Acknowledgments} We acknowledge insightful discussions with George Jackeli, Kyle Shen, Jaejun Yu.
This work was supported by NSF Grant DMREF-12-33349. 

\textbf{Note added:} While this manuscript was in preparation,
we became aware of work by Fujiyama \textit{et.al}~\cite{Fujiyama:2013},
where the ratio $\mu_L/\mu_S$ of 214 was estimated to be $2.5 \pm 0.35 $
by NRMXS measurements,
close to our theoretical value of $\mu_L/\mu_S\!\approx\! 2.2$.

\end{document}